\documentclass[twocolumn,twocolappendix]{aastex631}
\shortauthors{Kong et al.}
\shorttitle{}

\usepackage{booktabs}
\usepackage{multirow}
\begin{document}






\title{Cooler Phases of the Circumgalactic Medium Are More Centrally Concentrated: Constraints from Multiphase Absorption Lines}

\author[0009-0000-2569-2742]{Weiwen Kong}
\affiliation{School of Mechanics and Photoelectric Physics, Anhui University of Science and Technology, Huainan 232001, China}
\affiliation{Department of Astronomy, University of Science and Technology of China, Hefei 230026, China}
\affiliation{School of Astronomy and Space Science, University of Science and Technology of China, Hefei 230026, China}

\author[0009-0004-5989-6005]{Zeyu Chen}\thanks{Email:  \href{mailto:czy664@mail.ustc.edu.cn}{czy664@mail.ustc.edu.cn}}
\affiliation{Department of Astronomy, University of Science and Technology of China, Hefei 230026, China}
\affiliation{School of Astronomy and Space Science, University of Science and Technology of China, Hefei 230026, China}

\author[0000-0003-1588-9394]{Enci Wang}\thanks{Email:  \href{mailto:ecwang16@ustc.edu.cn}{ecwang16@ustc.edu.cn}}
\affiliation{Department of Astronomy, University of Science and Technology of China, Hefei 230026, China}
\affiliation{School of Astronomy and Space Science, University of Science and Technology of China, Hefei 230026, China}

\author[0009-0008-1319-498X]{Haoran Yu}
\affiliation{Department of Astronomy, University of Science and Technology of China, Hefei 230026, China}
\affiliation{School of Astronomy and Space Science, University of Science and Technology of China, Hefei 230026, China}

\author[0000-0002-3775-0484]{Kai Wang}
\affiliation{Institute for Computational Cosmology, Department of Physics, Durham University, South Road, Durham DH1 3LE, UK}
\affiliation{Centre for Extragalactic Astronomy, Department of Physics, Durham University, South Road, Durham DH1 3LE, UK}

\author[0000-0002-4314-5686]{Dongdong Shi}
\affiliation{School of Mechanics and Photoelectric Physics, Anhui University of Science and Technology, Huainan 232001, China}

\author[0009-0000-7307-6362]{Cheqiu Lyu}
 \affiliation{Department of Astronomy, University of Science and Technology of China, Hefei 230026, China}
 \affiliation{School of Astronomy and Space Science, University of Science and Technology of China, Hefei 230026, China}

 \author{Yuxuan Zhang}
 \affiliation{Department of Astronomy, University of Science and Technology of China, Hefei 230026, China}
 \affiliation{School of Astronomy and Space Science, University of Science and Technology of China, Hefei 230026, China}

 \author{Haoyi Zhang}
 \affiliation{Department of Astronomy, University of Science and Technology of China, Hefei 230026, China}
 \affiliation{School of Astronomy and Space Science, University of Science and Technology of China, Hefei 230026, China}

 \author{Haowen Guan}
 \affiliation{Department of Astronomy, University of Science and Technology of China, Hefei 230026, China}
 \affiliation{School of Astronomy and Space Science, University of Science and Technology of China, Hefei 230026, China}

\begin{abstract}

\noindent {We present a systematic study of the multiphase circumgalactic medium (CGM) around galaxies and quasars, traced by Ca\,\textsc{ii}\,$\lambda\lambda$3934,\,3969, Mg\,\textsc{ii}\,$\lambda\lambda$2796,\,2803, and C\,\textsc{iv}\,$\lambda\lambda$1548,\,1550, using the Year~1 dataset from the Dark Energy Spectroscopic Instrument.  These three doublets trace CGM gas across a range of temperatures, from cold to warm phases, and we employ a stacking technique to measure the corresponding absorption signals using background sources.  
We show that CGM structure is strongly phase-dependent: ions tracing progressively cooler gas exhibit increasingly steep radial profiles in equivalent width of absorption ($W_i$). These trends are broadly consistent with predictions from cosmological simulations, supporting a phase-stratified CGM in which cooler gas is more centrally concentrated.
Specifically, halos of emission-line galaxies exhibit a strong radial transition from cool to warm gas, whereas halos of quasars show a more uniform distribution, likely regulated by active galactic nuclei feedback; in contrast, the cold gas traced by Ca\,\textsc{ii} in low-redshift galaxies is tightly confined to inner regions.
We further demonstrate that the radial scaling $W_i \propto D^{\alpha}$ is primarily set by host stellar mass, particularly for the cool-phase medium, suggesting efficient heating processes in massive halos. 
By jointly leveraging multiple absorption tracers from observation and simulations, we map the CGM from cold to warm phases and place new constraints on the baryon cycle that governs galaxy evolution.
}

\end{abstract}



\section{Introduction} \label{sec:intro}
The circumgalactic medium (CGM) serves as the critical interface between galaxies and the surrounding large-scale structure, playing a fundamental role in regulating galaxy formation and evolution~\citep[e.g.,][]{tumlinson17,chen26cgm}. This surrounding medium regulates the supply of gas for star formation, the redistribution of metals, and the impact of feedback processes, while also governing gas accretion, outflow recycling, and the thermal and dynamical state of baryons in galaxy halos~\citep{peroux2020,faucher2023, Yu2025,Lyu2026}. Thus, observations of the CGM, together with their connection to galaxy properties, provide key constraints on models of galaxy evolution, especially on the subgrid prescriptions of stellar and black hole feedback~\citep{ford16,angls17,wang2019,wang22origin,wang22gas, Ma2024, Jia2025, Lyu-25, peroux2024,Wang2026}.

It is widely recognized that the CGM is a highly multiphase and dynamically complex medium, spanning a wide range of temperatures and densities, from cool ($T \sim 10^4$\,K) dense clouds to hot ($T \gtrsim 10^6$\,K) diffuse halo gas~\citep[e.g.,][]{werner14,zahedy19,olivares22,dutta24}. Numerical simulations indicate that cold-mode accretion~\citep{dekel2009}, feedback-driven outflows~\citep{shah25}, and gas recycling processes coexist and interact within the CGM, giving rise to filamentary structures, turbulent mixing layers, and condensation driven by thermal instabilities~\citep{hafen20,Hafen-22,weng24}. These different phases in the CGM exhibit substantial differences in their spatial distribution and evolutionary behavior~\citep{faucher2023,saeedzadeh23}.

EAGLE zoom simulations show that low-ionization metal gas predominantly resides in cool, dense clouds in the inner CGM, while O\,\textsc{vi} traces a spatially and physically distinct, diffuse high-temperature phase~\citep{oppenheimer18}. 
The {\tt Eris} simulation also highlights that hot, metal-enriched gas ($T > 10^5\,\mathrm{K}$) is preferentially transported into the outer CGM through high-velocity outflows~\citep{shen12}. 
Cool, photoionized gas ($T \sim 10^4\,\mathrm{K}$), commonly traced by C\,\textsc{ii} and Si\,\textsc{ii}, exhibits a rapidly declining covering fraction beyond the virial radius ($R_{\mathrm{vir}}$). In contrast, high-ionization species such as O\,\textsc{vi} \citep[e.g.,][]{sanchez19,li21} and C\,\textsc{iv} maintain substantial covering fractions out to large distances, with O\,\textsc{vi}-enriched halos extending to $\sim 4\,R_{\mathrm{vir}}$ \citep{shen13}.

Although simulations suggest a well-defined double-exponential structure of the CGM, with an inner region dominated by dense, low-ionization gas and an outer region dominated by diffuse, high-ionization gas~\citep{liang16}, observationally probing this multiphase nature of the CGM remains extremely challenging due to its intrinsically low surface brightness~\citep{lokhorst19,zaritsky19}. As a result, absorption-line studies that employ phase-sensitive tracers are essential for effectively characterizing its different components~\citep{zhu14,liang14,Huang16,ng2025,chen25b}. For example, Mg\,\textsc{ii} and Ca\,\textsc{ii} lines are commonly used to trace cool gas ($\sim 10^{4}\,\mathrm{K}$)~\citep[e.g.,][]{bergeron86,chen10absorp,chen10cool,zhu13ca,lan18,anand21,chen25,ng2025}, while C\,\textsc{iv} ($\sim 10^{4.5}$--$10^{5}\,\mathrm{K}$) and O\,\textsc{vi} ($\sim 10^{5.5}\,\mathrm{K}$) trace relatively warmer gas~\citep[e.g.,][]{tumlinson11,Bordoloi14c,Kacprzak15,garza25}.

Based on multi-ion observations of individual absorption systems (e.g., H\,\textsc{i}, Mg\,\textsc{ii}, Si\,\textsc{ii}, C\,\textsc{iv}, and O\,\textsc{vi}), \cite{Sameer24} employ a cloud-by-cloud multiphase ionization modeling approach and demonstrate that the CGM is not a continuous medium but is instead composed of a superposition of numerous, highly fragmented discrete clouds spanning cool ($\sim10^{4}\,\mathrm{K}$), warm ($\sim10^{4.5}$--$10^{5}\,\mathrm{K}$), and hot ($\sim10^{5.5}$--$10^{6}\,\mathrm{K}$) phases. This means that multiphase absorption features can overlap along the same line of sight (LOS), but typically do not arise from the same physical gas cloud~\citep{werk16}. On global scales, \cite{lan25} showed from individual absorber measurements that Mg\,\textsc{ii} predominantly traces gas in the inner CGM, close to galaxies, whereas C\,\textsc{iv} extends to larger radii, reaching the outer CGM and extending into the intergalactic medium (IGM). Kinematically, quasar sightline observations near the galaxy major axis~\citep[e.g.,][]{DeFelippis21,Kacprzak25,Ho26} show that low-ionization gas (e.g., H\,\textsc{i}, Mg\,\textsc{ii}, and Si\,\textsc{ii}) exhibits stronger co-rotation with the host galaxy disk and more ordered motion, whereas high-ionization gas (e.g., O\,\textsc{vi}) shows weaker co-rotation and more kinematically diffuse behavior.

As multi-wavelength observations are required to probe different tracers~\citep{tumlinson17,nelson2025}, achieving a consistent characterization of the multiphase CGM remains highly challenging. Despite substantial progress in previous studies, a unified view of the CGM across temperature regimes remains limited by the reliance on either small samples of individually detected absorbers~\citep{lan17,anand21,Zou-24a,Zou-24b} or single-phase analyses~\citep[e.g.,][]{zhu13ca,zhu14,wu2024,ng2025}. Thus, with the goal of establishing a statistically robust connection between different ionic tracers and fully capturing the ensemble properties of multiphase gas in galaxy halos, this work leverages statistical samples of thousands of galaxy--quasar (or quasar--quasar) pairs and employs a stacking approach that preserves the ensemble properties of the CGM while mitigating biases associated with independent detections of individual absorption lines. 

We utilize the Year~1 (Y1) dataset from the Dark Energy Spectroscopic Instrument (DESI) to systematically examine three representative absorption features tracing multiple CGM gas phases from cold to warm: the Ca\,\textsc{ii}\,H\,\&\,K\, $\lambda\lambda$3934,\,3969 ($\lesssim 10^4\,$K), Mg\,\textsc{ii}\,$\lambda\lambda$2796,\,2803 ($\sim10^4\,$K), and C\,\textsc{iv}\,$\lambda\lambda$1548,\,1550 ($\sim10^{4.5}$–$10^{5.5}\,$K) doublets.
Furthermore, we compare our results with DESI-like synthetic spectra from TNG50 provided by the Synthetic Absorption Line Spectral Almanac (SALSA; \citealt{nelson2025}), offering a physically motivated framework for interpreting the observed multiphase CGM properties.
This work is structured as follows. We introduce the DESI Y1 data, sample selection, and spectral stacking methodology in Section~\ref{sec:data}. Section~\ref{sec:result} presents our key findings: the spatial distribution of the multiphase gas traced by the three absorption lines and the comparisons with simulations, as well as the dependence of their radial profiles on host galaxy redshift and stellar mass. In Section~\ref{sec:summary}, we summarize the implications for the multiphase CGM and present future prospects.
We adopt a flat $\Lambda$CDM cosmology with $h=0.677$,  $\Omega_{\rm M}=0.309$ and $\Omega_{\rm \Lambda}=0.691$ using the \texttt{Planck15}\footnote{\url{https://docs.astropy.org/en/latest/api/astropy.cosmology.realizations.Planck15.html}} package \citep{2016A&A...594A..13P}.

\section{Data analysis} \label{sec:data}

\subsection{DESI Y1 Dataset} \label{subsec:DESI}
DESI is a Stage-IV ground-based dark energy experiment~\citep{levi19} installed on the Mayall 4-meter telescope at Kitt Peak National Observatory. Equipped with 5,000 robotic fiber positioners and a wide field of view, DESI is capable of obtaining spectra for thousands of galaxies and quasars simultaneously in a single exposure~\citep{Silber23,miller24,poppett24}. The instrument's high throughput and spectral resolution ($R \sim 2000–5100$) over a broad wavelength coverage ($3600–9824\,\rm \AA$) make it an ideal facility for investigating the diffuse CGM through absorption line spectroscopy~\citep[e.g.,][]{chen25,chen25b,lan25}.
Extragalactic sources in DESI can be categorized into four classes~\citep{myers23}, including QSOs ($0.2<z < 6.0$; \citealt{yeche20,chaussidon23}) and three types of galaxies: Bright Galaxy Survey galaxies (BGSs; $0 <z < 0.6$; \citealt{ruiz20,hahn23}), Luminous Red Galaxies (LRGs; $0.3 < z < 1.2$; \citealt{zhou20,zhou23}), Emission Line Galaxies (ELGs; $0.6 < z < 1.6$) \citep{raichoor20,raichoor23}. 

DESI Y1, as the first major data release of the DESI collaboration, contains spectral data for more than 18 million targets~\citep{DESI2025}. This release represents a significant leap in statistical power compared to the Early Data Release (EDR), which was published in June 2023 \citep{adame23}, offering an order-of-magnitude increase in the number of extragalactic spectra. This large dataset enables unprecedented statistical sensitivity to detect weak metal absorption lines and to probe multiphase gas across diverse halo environments~\citep[e.g.,][]{wu2024,ng2025}.



\subsection{Pair Selection and Sample Properties} \label{subsec:pair}

\begin{figure*}
    \centering
    \includegraphics[width=1.8\columnwidth]{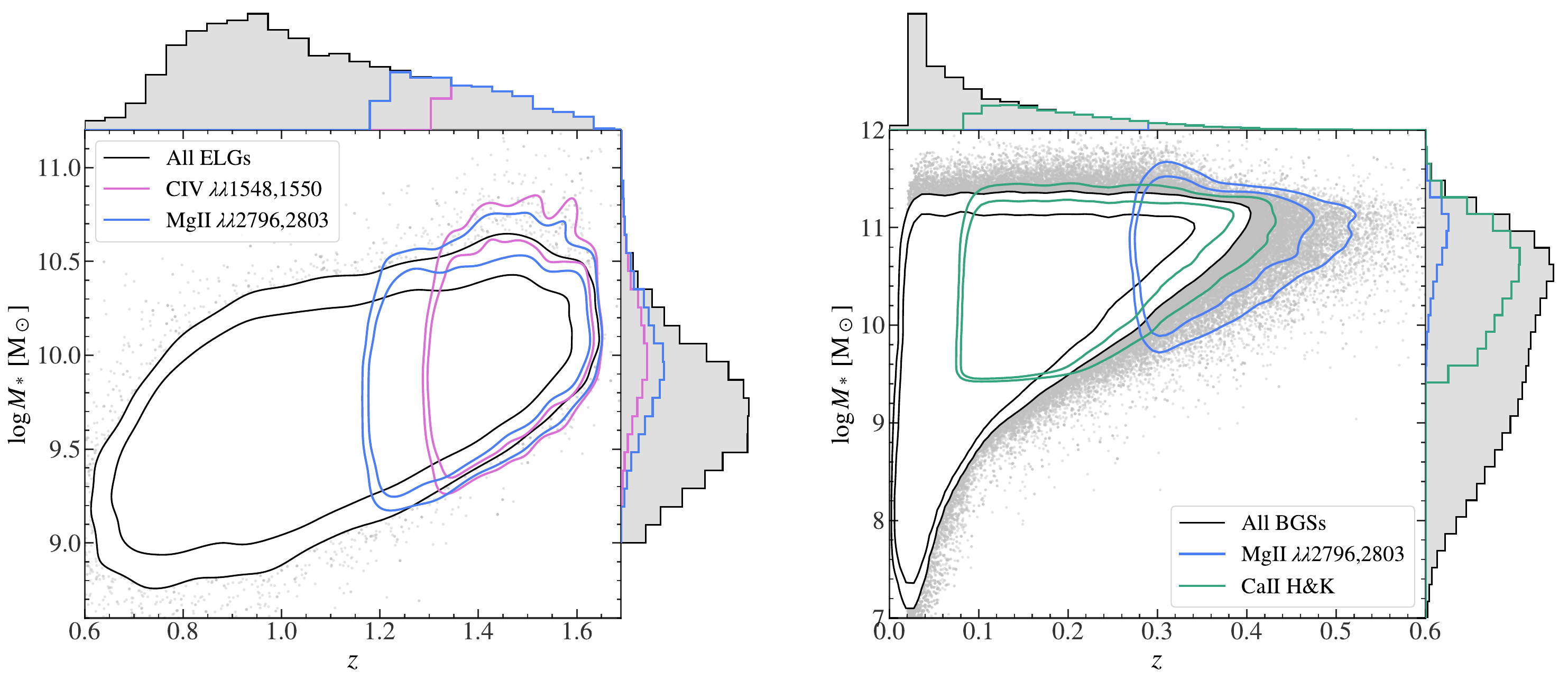}
    \caption{Stellar mass–redshift distributions of the foreground galaxy samples used for the absorption-line stacking analysis. The left panel shows ELGs, and the right panel shows BGSs. Gray points represent the full parent samples, with black contours indicating the 1-$\sigma$ and 2-$\sigma$ density levels. Colored contours denote the subsamples with spectral coverage suitable for different absorption-line tracers: Mg\,\textsc{ii} (blue; $z>1.2$ for ELGs and $z>0.3$ for BGSs), C\,\textsc{iv} (magenta; $z>1.3$; ELGs only), and Ca\,\textsc{ii} (green; $z>0.1$ and $\log M_*/{\rm M}_{\odot}>9.4$; BGSs only). The marginal histograms show the corresponding one-dimensional distributions in redshift and stellar mass for the parent samples (gray) and the selected subsamples (colored). These subsamples define the foreground–background pairs used to probe multiphase CGM through Mg\,\textsc{ii}–C\,\textsc{iv} (ELGs) and Ca\,\textsc{ii}–Mg\,\textsc{ii} (BGSs) comparisons.}
    \label{fig_pair}
\end{figure*}

\begin{table*}[ht]

\centering

\caption{Summary of foreground samples and subsamples within $D<150\,$kpc.}

\label{tab:sample_summary}

\begin{tabular}{llcl}

\toprule

Sample &
Selection &
$N_{\rm pair}$ &
Used for \\

\midrule

\multirow{3}{*}{ELG}

& Full sample
& 33072
& Reference sample \\

&$z>1.2$
& 7735
& Mg\,\textsc{ii} \\

&$z>1.3$
& 6670
& C\,\textsc{iv} \\

\midrule

\multirow{3}{*}{BGS}

& Full sample
& 1044958
& Reference sample \\

& Mg\,\textsc{ii}: $z>0.3$,
& 119101
& Mg\,\textsc{ii} \\

& Ca\,\textsc{ii}: $z>0.1$, $\log M_*/M_\odot>9.4$
& 424263
& Ca\,\textsc{ii}\\

\midrule

\multirow{2}{*}{QSO}

&Full sample
& 19601
& Reference sample \\

&$z>1.3$
& 5979
& Mg\,\textsc{ii}$+$C\,\textsc{iv} \\

\bottomrule

\end{tabular}

\end{table*}

We construct our absorption-line sample by cross-matching foreground ELGs, BGSs, and QSOs with background quasars in the DESI Y1 dataset. Here, galaxies and QSOs are distinguished based on the \texttt{SPECTYPE} classification from the \texttt{zcatalog}\footnote{\url{https://data.desi.lbl.gov/public/dr1/spectro/redux/iron/zcatalog/v1/zall-tilecumulative-iron.fits}}. Redshifts are taken from the same catalog and are derived from spectroscopic data using the automated \texttt{redrock} pipeline\footnote{\url{https://github.com/desihub/redrock}} \citep{anand24}. To ensure the reliability of the redshift measurements, we select sources with \texttt{ZWARN} = 0~\citep{schlafly23}. The stellar masses ($M_*$) of BGS and ELG samples are estimated via spectral energy distribution (SED) modeling with \texttt{CIGALE} \citep{boquien19}. The modeling is based on $g$, $r$, $z$, WISE1, and WISE2 photometry extracted using the \texttt{Tractor} algorithm \citep{lang16}, as provided in the DESI Mass-EMLines catalog\footnote{\url{https://data.desi.lbl.gov/doc/releases/dr1/vac/stellar-mass-emline/}} \citep{zou24}.

We use background sightlines that pierce through the halos of foreground objects at projected distances ($D$) ranging from 10 to 150 kpc (for ELGs and BGSs) and 15 to 150 kpc (for foreground QSOs). The lower limits on $D$ are adopted to minimize the impact of the intrinsic radiation fields of the foreground sources~\citep{lan18,chen25b}. The upper limit of 150\,kpc approximately corresponds to the typical virial radius ($R_{\mathrm{vir}}$) of our ELG sample with $M_* \sim 10^{10}\,\rm M_\odot$~\citep{bryan98}, thereby ensuring that the analysis predominantly probes the CGM within the host dark matter halo.

To ensure that the detected absorption arises from intervening halos rather than the local environment of the background quasars, we require a velocity separation of $v_{\rm off} > 6000\,\mathrm{km/s}$~\citep{napolitano23}, where $v_{\rm off} =c\times (z_{\rm bg} - z_{\rm fg})/(1 + z_{\rm fg})$ is defined from the redshift difference between the foreground ($z_{\rm fg}$) and background ($z_{\rm bg}$) objects. 
Because Ly$\alpha$ absorption does not introduce coherent biases in absorption-line measurements as demonstrated in Appendix~B of \cite{chen25b}, we do not explicitly exclude galaxy–QSO pairs—even when the absorption features fall within the Ly$\alpha$ forest of the background QSO \citep{Pieri14, Morrison24}, i.e., $\lambda_{i}^{\mathrm{rest}} (1+z_{\mathrm{fg}}) < 1216\,\text{\AA}\times(1+z_{\mathrm{bg}})$, where $\lambda_{i}^{\mathrm{rest}}$ is the rest-frame wavelength of the transition. 
\begin{figure*}
    \centering
    \includegraphics[width=1.8\columnwidth]{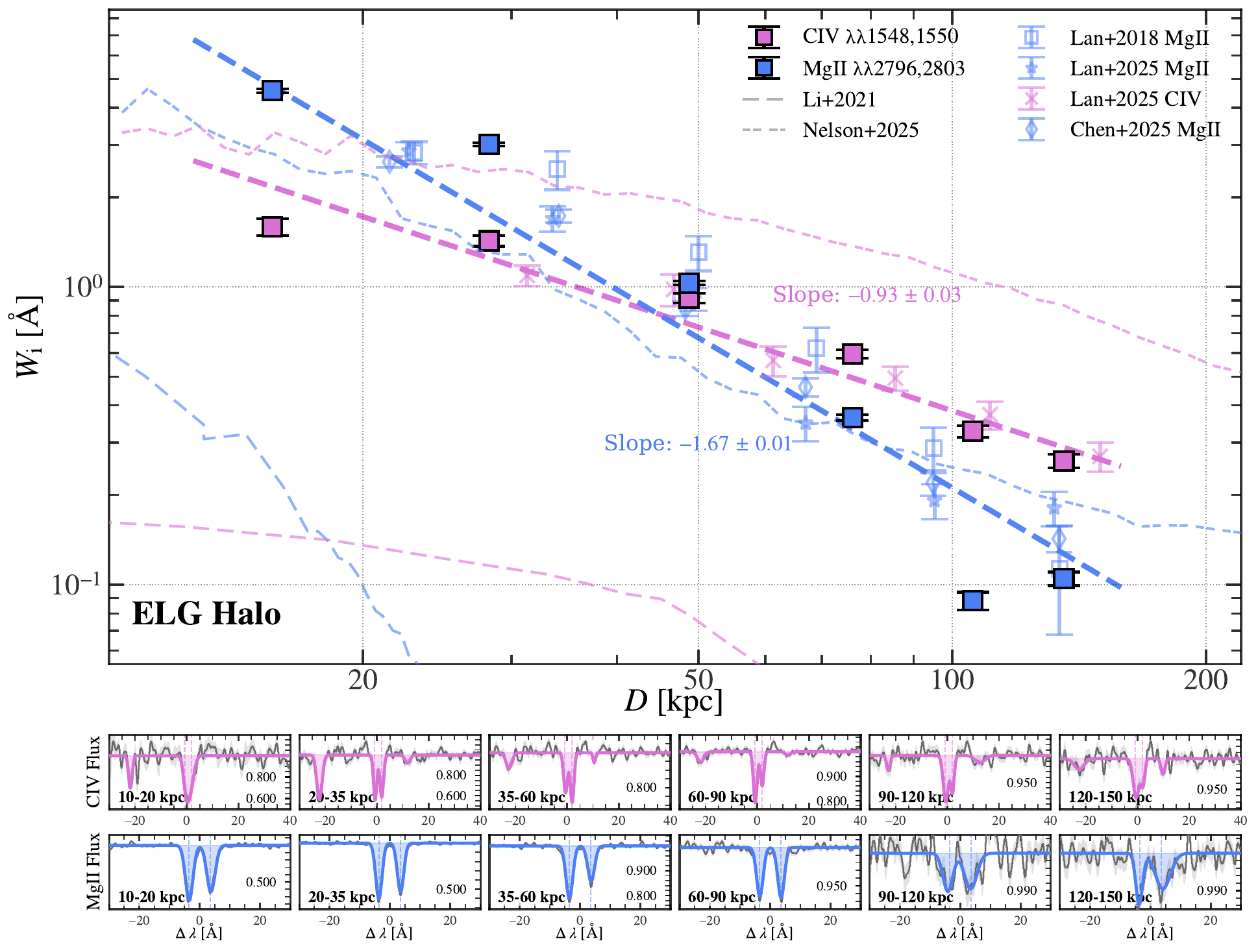}
    \caption{The radial profiles and corresponding stacked spectra of representative CGM absorption lines around ELG host galaxies selected from the left panel of Figure~\ref{fig_pair}. The colors denote the corresponding absorption lines: C\,\textsc{iv} (pink) and Mg\,\textsc{ii} (blue). Top panel: The $W_i$ as a function of $D$ for C\,\textsc{iv} $\lambda\lambda$1548, 1550 (pink squares) and Mg\,\textsc{ii} $\lambda\lambda$2796, 2803 (blue squares). Error bars denote the 1-$\sigma$ uncertainties derived from RMS of the flux errors within the 3-$\sigma$ wavelength range of the Gaussian fit shown in the lower panels. The dashed lines represent the best-fit power-law relations, $W_i \propto D^{\alpha}$, with the corresponding slope $\alpha$ and its uncertainty labeled in the panel. For comparison, the open symbols show observational results from previous studies, including stacked spectra analyses~\citep{lan18,chen25} and individual absorber measurements~\citep{lan25}. The dashed and dash–dash pattern curves represent the median $W_i$ profile from the FIRE simulations~\citep{li21} and the mean $W_i$ profile from TNG50~\citep{nelson2025}, respectively, both corresponding to galaxies with stellar masses of $\sim 10^{10}\, \rm {M_\odot}$.  Bottom panels: The normalized stacked spectra are shown in the rest-frame wavelengths of the foreground galaxies, with $\Delta \lambda$ centered at 2800\,\AA\ (Mg\,\textsc{ii}) and 1549\,\AA\ (C\,\textsc{iv}), respectively.  The black curves show the stacked spectra, while the colored curves show the corresponding Gaussian fits. The shaded gray regions indicate the 1-$\sigma$ uncertainties, and the colored shaded areas correspond to the measured equivalent widths.  A local fluctuation is present around the $\sim100$ kpc bin in the Mg\,\textsc{ii} profile, which may arise from sample variance, statistical fluctuations, or residual systematic uncertainties. However, it does not significantly affect the power-law fit or the overall radial trend.}
    \label{fig:EW_spec_ELG}
\end{figure*}

\subsection{Spectral Processing and Stacking Method} \label{subsec:stacking}

The spectral processing and stacking methodology adopted in this work largely follows the robust pipeline developed in \cite{chen25,chen25b}. We first perform continuum fitting on each background QSO spectrum using the \texttt{PyQSOFit} code \citep{guo18} to remove the intrinsic emission features. After normalizing the spectra by the fitted continuum, we apply a double-iteration median filter to mitigate residual large-scale fluctuations. A crucial step is to mask the expected wavelength regions of the target absorption lines (i.e., Ca\,\textsc{ii}, Mg\,\textsc{ii}, and C\,\textsc{iv}) prior to filtering. In the first iteration, we use a window size of 141 pixels ($\sim113$ \AA) to capture continuum trends, followed by a finer window of 71 pixels ($\sim57$ \AA) in the second iteration~\citep[see also][]{zhu13,zhu14,lan18}. To prevent the suppression of genuine absorption signals, we mask the spectral regions corresponding to expected metal lines in both iterations, spanning $\sim4$ $\rm {\AA}$ centered on the absorption line cores. This step ensures that the absorption troughs are preserved while the local continuum is effectively flattened.

We then shift the processed spectra to the rest frame of the foreground targets using $z_{\rm fg}$. Since the observation-frame pixels do not align in the rest frame, we re-grid all spectra onto a uniform wavelength array spanning 900\,\AA\ with a fixed pixel scale of 0.2\,\AA, corresponding to the native DESI spectral resolution when shifted to the highest-redshift quasars in our sample, thereby minimizing information loss in individual spectra, albeit at a modest increase in computational cost. At each wavelength pixel, we compute the median flux of all available spectra in the stack to obtain the final median composite spectrum. The signal-to-noise ratio (S/N) of the composite spectrum increases steadily with sample size and scales as $N_{\mathrm{spec}}^{0.5}$ as expected \citep{chen25}, where $N_{\mathrm{spec}}$ is the number of spectra included in the stack. As an unbiased approach, this median stacking is widely adopted due to its flexibility and robustness~\citep[e.g.,][]{zhu13,zhu14,lan18,chen25b,wu2024,ng2025,Yu2025}.

\subsection{Measurement} \label{subsec:measure}

We estimate the uncertainty of the composite spectrum at each wavelength pixel using a bootstrap resampling approach with 100 realizations, which is sufficient for convergence. To quantify the absorption strength, we fit each absorption feature with (double) Gaussian profiles using the \texttt{emcee} sampler\footnote{\url{https://emcee.readthedocs.io/en/stable/index.html}} \citep{foreman13} and calculate the rest-frame equivalent width ($W_i$) by integrating the best-fit Gaussian models over a window of $\pm\,3\,\sigma$ from the line centroid. The measured $W_i$, along with the mean redshift and stellar mass of the galaxy sample in each bin, are cataloged for the subsequent analysis of scaling relations. The uncertainty of the $W_i$ is estimated as the root mean square (RMS) of the flux errors within the same wavelength interval~\citep[see][for details]{chen25}.

\section{RESULTS} \label{sec:result}


\begin{figure*}
    \centering
    \includegraphics[width=1.8\columnwidth]{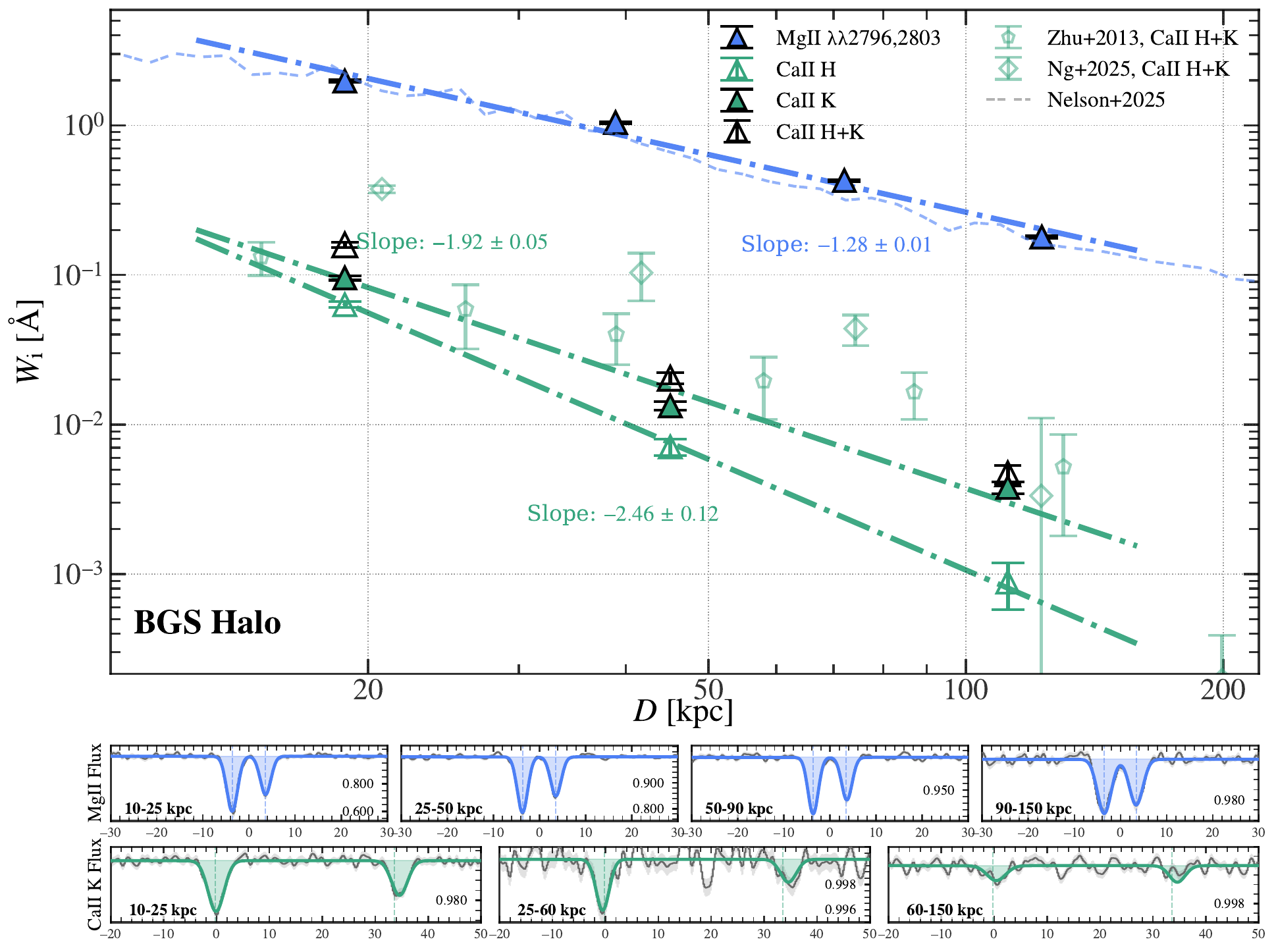}
    \caption{Same as Figure~\ref{fig:EW_spec_ELG}, but for the BGS sample selected from the right panel of Figure~\ref{fig_pair}. The Ca\,\textsc{ii} absorption is shown in green.  The top panel shows Mg\,\textsc{ii} $\lambda\lambda$2796, 2803 (blue triangles) and Ca\,\textsc{ii}\,H\,\&\,K (green open and filled triangles, respectively). The pentagon and diamond symbols and lines show results from previous studies by~\cite{zhu13ca} (ELGs, $M_* \sim 10^{10.3}\,\rm M_\odot$) and~\cite{ng2025} (BGSs, $M_* \sim 10^{10.7}\,\rm M_\odot$), respectively. We show the Ca\,\textsc{ii} H and K lines separately as a consistency check of the doublet behaviour, and additionally present their summed equivalent width (black open triangles) to facilitate direct comparison. We also present the $W$ profile from SALSA with stellar mass ($10 < \log (M_*/M_{\odot}) < 11.2$) and redshift ($z=0.5$) most closely matching those of the BGS sample used in our Mg\,\textsc{ii} stacking analysis as a reference \citep{nelson2025}. Bottom panels: similar to Figure~\ref{fig:EW_spec_ELG}, $\Delta \lambda$ for Ca\,\textsc{ii} is centered at 3935\,\AA.}
    \label{fig:EW_spec_BGS}
\end{figure*}

To investigate the multiphase nature of the CGM, we focus on combinations of absorption lines that trace different gas phases. Taking C\,\textsc{iv} as an example, a redshift of $z \gtrsim 1.4$ is typically required for the line to be redshifted into the optical wavelength range, thereby enabling ground-based observations~\citep{turner14,Dutta-21,galbiati23}. For the DESI wavelength coverage of 3600–9824\,\AA, the accessible redshift ranges of several commonly used CGM tracers can be derived from the condition $3600\,\text{\AA} < \lambda^{\rm rest}_i (1+z_{\rm fg}) < 9824\,\text{\AA}$, which yields the following redshift ranges: Ca\,\textsc{ii} ($0 < z < 1.5$), Mg\,\textsc{ii} ($0.3 < z < 2.5$), C\,\textsc{iv} ($1.3 < z < 5.3$), and O\,\textsc{vi} ($2.5 < z < 8.5$).

Therefore, for higher-redshift ELG and foreground QSO halos, 
we focus on the high-ionization C\,\textsc{iv} doublet and the low-ionization Mg\,\textsc{ii} doublet, to trace distributions of the warm and cool medium. 
For lower-redshift BGSs, where C\,\textsc{iv} falls outside the blue end of the observable wavelength range, we instead compare the cold-phase Ca\,\textsc{ii} H\,\&\,K lines with Mg\,\textsc{ii} doublet. Studying O\,\textsc{vi} requires relatively high redshifts ($z > 2.5$), where only a limited number of foreground QSOs are available, making it difficult to assemble sufficiently large pair samples for stacking to achieve detectable signals. As a result, the investigation of hotter gas phases ($T \gtrsim 10^{5.5}\,\mathrm{K}$) is beyond the scope of this work.

\subsection{The Spatial Distribution of Multiphase Gas} \label{subsec:ew}


\begin{figure*}
    \centering
    \includegraphics[width=1.8\columnwidth]{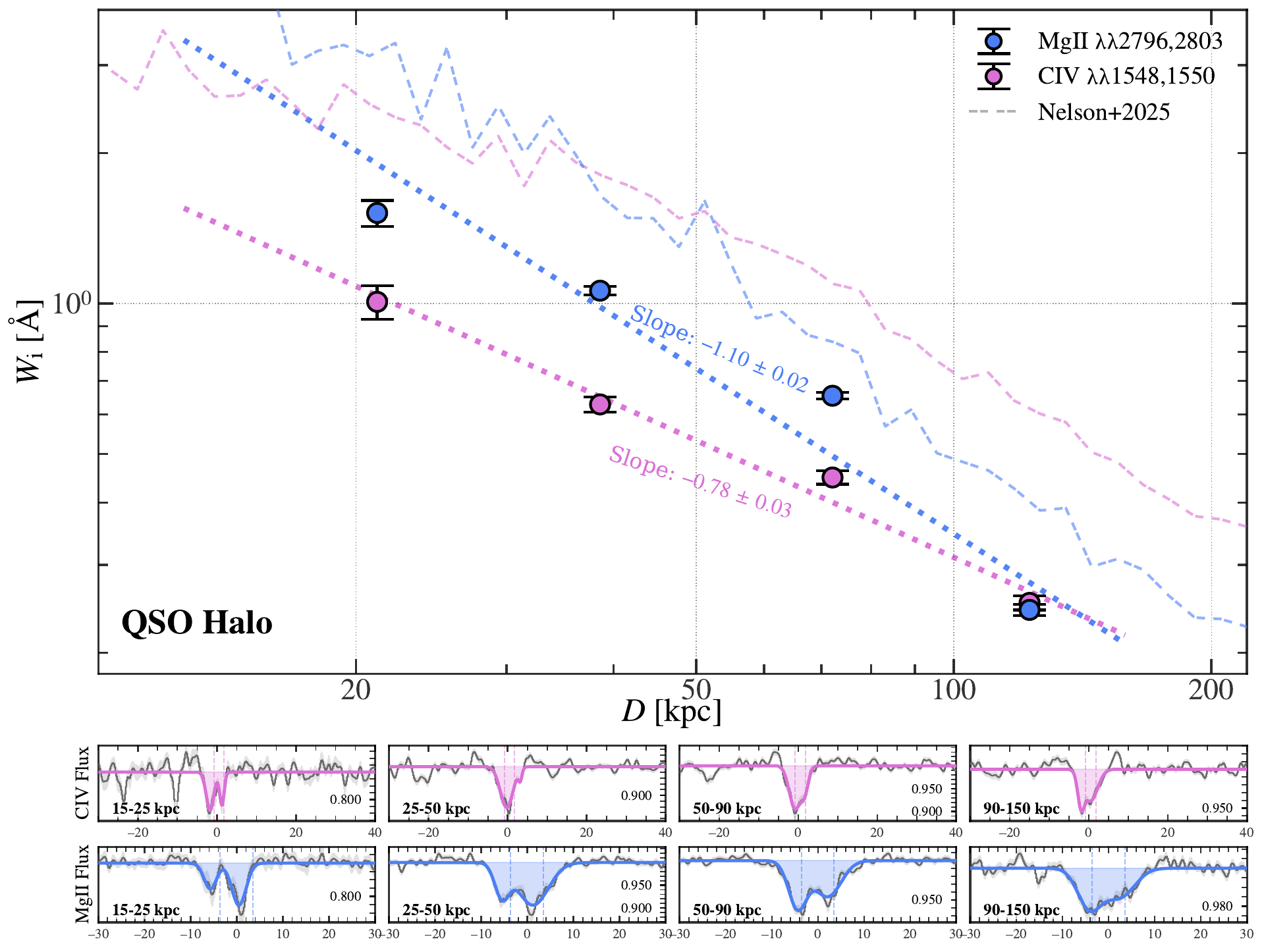}
    \caption{Same as Figure~\ref{fig:EW_spec_ELG}, but for the QSO sample with overlapping C\,\textsc{iv} and Mg\,\textsc{ii} coverage ($z>1.3$). The top panel shows C\,\textsc{iv} $\lambda\lambda$1548, 1550 (pink circles) and Mg\,\textsc{ii} $\lambda\lambda$2796, 2803 (blue circles), while the bottom panels show the stacked spectra corresponding to each data point. The curves represent results from SALSA for halos with $12 < \log (M_{\rm halo}/\rm {M_{\odot}}) < 13$ at $z=2$ \citep{nelson2025}, which is close to the mean redshift of our QSO sample and broadly consistent with the typical halo mass range of QSO host halos~\citep{zhao13,Timlin18}.}
    \label{fig:EW_spec_QSO}
\end{figure*}

We first focus on the differences in the radial profiles of the equivalent widths of the three absorption lines. Although the absolute equivalent widths are not directly comparable across tracers, their radial profiles, and in particular their gradients with impact parameter, provide a robust probe of differences in the spatial distribution of different CGM phases. To enable a controlled comparison, we construct foreground galaxy samples with well-defined and consistent properties. Figure~\ref{fig_pair} shows the stellar mass--redshift distributions of the ELG and BGS samples used in this work. Across the full galaxy sample, ELGs have a median stellar mass of $\sim10^{9.6}\,\rm M_{\odot}$, which increases mildly with redshift due to sample selection, while BGSs are more massive, with a median stellar mass of $\sim10^{10}\,\rm M_{\odot}$, and exhibit a clear observational selection limit.

The two populations are separated in redshift, which naturally provides complementary access to different CGM tracers. For ELGs (left panel of Figure~\ref{fig_pair}), we select galaxies at $z>1.2$, where over 85\% of the spectra simultaneously cover both C\,\textsc{iv} and Mg\,\textsc{ii}, enabling a consistent stacking analysis in which the two ions trace the same CGM population. For the BGS sample (right panel), the weaker Ca\,\textsc{ii} signal necessitates extending the sample to $z\sim 0.1$ in order to achieve sufficient S/N.
This leads to only $\sim$30\% overlap between the Mg\,\textsc{ii} ($z>0.3$) and Ca\,\textsc{ii} ($z>0.1$) samples. Nevertheless, by excluding low-redshift, low-mass galaxies ($\log M_*/{\rm M}_{\odot}>9.4$), we ensure that the resulting subsamples remain comparable in stellar mass ($\sim 10^{10.7}\,M_{\odot}$). Table~\ref{tab:sample_summary} further summarizes the sample size ($N_{\rm pair}$) and selection shown in Figure~\ref{fig_pair}, as well as the number of quasar-quasar pairs used in the QSO foreground analysis. These sample sizes are sufficiently large to enable high-quality stacked spectra within $D < 150$\,kpc.


Figure~\ref{fig:EW_spec_ELG} presents the radial profiles of $W_{\rm C\,IV}$ and $W_{\rm Mg\,II}$ around ELGs derived from the high-signal-to-noise stacked spectra shown in the bottom panels, corresponding to the foreground sample shown in the left panel of Figure~\ref{fig_pair}. As shown, these absorption signals are highly robust. Our measurements of $W_{\rm Mg\,\textsc{ii}}$ around ELGs are consistent with previous stacked analyses based on the Sloan Digital Sky Survey (SDSS; \citealt{menard11})~\citep{lan18} and the DESI EDR~\citep{chen25}. The radial profile of $W_i$ is well described by a power-law relation, $W_i \propto D^{\alpha}$, indicating a smooth monotonic decline of Mg\,\textsc{ii} absorption strength with projected distance. The warm gas traced by C\,\textsc{iv} exhibits a relatively shallow slope of $-0.93 \pm 0.03$, indicating a diffuse and extended envelope. In contrast, the cool gas traced by Mg\,\textsc{ii} is more centrally concentrated, characterized by a steeper slope of $-1.67 \pm 0.01$. These measurements are consistent with the mean $W$ values reported in \cite{lan25}, which include both detected and non-detected sources. From their results, we obtain $\alpha = -1.73$ for Mg\,\textsc{ii} and $\alpha = -0.92$ for C\,\textsc{iv} over $20 < D < 150$ kpc and $10 < \log M_\ast < 10.4$. 

We also compare the results with synthetic absorption measurements from simulations of galaxies with similar stellar masses ($\sim 10^{10}\,\rm M_{\odot}$). However, we note that the limited resolution and incomplete treatment of radiation fields in simulations may lead to equivalent widths that are up to $\sim 1$ dex lower than observations~\citep[e.g.,][]{li21} and produce overly steep column density radial profiles~\citep[e.g.,][]{nelson20,cook25}. Nevertheless, the simulations exhibit broadly consistent behavior across different ions, with a general trend in which colder gas shows steeper radial profiles. As shown in Figure~\ref{fig:EW_spec_ELG}, both the FIRE simulations (based on median equivalent widths; \citealt{li21}) and the SALSA~\citep{nelson2025} from TNG50 (based on mean equivalent widths) predict that C\,\textsc{iv} exhibits a more spatially extended and stable distribution than Mg\,\textsc{ii}. SALSA generates a large number of sightlines from different cosmological simulations: e.g., TNG50~\citep{pillepich19} and EAGLE~\citep{schaye15}, producing synthetic spectra that correspond to a variety of observational surveys: e.g., DESI, HST/COS~\citep{green2012}, and SDSS-BOSS~\citep{dawson13}. This framework allows us to select sightlines passing near foreground galaxies with properties matched to those in observations, enabling a direct comparison between simulated and observed absorption features. An illustrative example is presented in Appendix~\ref{sec:appenA}.

In simulations, the surrounding CGM actively interacts with host galaxies, comprising both inflowing and outflowing gas.
Feedback-driven outflows provide a natural physical explanation for these observed ion-dependent trends: low-ionization species trace relatively recent outflows, in which cool, dense gas has been recently launched from the galactic disk and has not yet propagated far before being slowed or recycled by gravity, forming a fountain-like cycle~\citep{liang16}. By contrast, high-ionization species are associated with older outflows that were expelled at earlier times, subsequently advected to larger radii and modified by long-term expansion and shock heating, reaching warmer phases that are more spatially extended. However, the inflow scenario also provides a plausible explanation for these ion-dependent trends. The low-ionization tracers (e.g., Mg\,\textsc{ii}, Si\,\textsc{ii}, C\,\textsc{ii}) probe cool, compact clumps that are likely in a recycling or inflowing phase, and are expected to accrete onto the galaxy on Gyr timescales, while high-ionization species trace a hot, diffuse halo medium \citep{oppenheimer18}. 
The inner regions of halos have higher baryon densities and therefore higher cooling rates, which can lead to a more concentrated distribution of cool gas~\citep{Sutherland93,ford16,Voit17}. In practice, the CGM likely reflects a combination of both effects.




Figure~\ref{fig:EW_spec_BGS} focuses on the low-redshift BGS sample (the right panel of Figure~\ref{fig_pair}), extending our analysis to the cooler gas phase traced by Ca\,\textsc{ii}. The radial profile of Mg\,\textsc{ii} (blue triangles) has a power-law slope of $-1.28 \pm 0.01$ and exhibits significantly stronger absorption than Ca\,\textsc{ii}. The Ca\,\textsc{ii} K line shows a rapid radial decline with a slope of $-1.92 \pm 0.05$. Although Ca\,\textsc{ii} signals can still be detected out to $\sim 150$ kpc, the absorption strength drops below 0.1 Å beyond 50 kpc. The Ca\,\textsc{ii} H line (open triangles) follows a similar trend but is systematically weaker than the K line, consistent with the expected 1:2 ratio of their oscillator strengths~\citep{kramida2024}. These behaviors are consistent with stacked measurements of the Ca\,\textsc{ii} doublet in SDSS ELGs~($\alpha = -1.38$\,; \citealt{zhu13ca}) and DESI Y1 BGSs~($\alpha = -2.15$\,; \citealt{ng2025}). The steeper slope of $W_{\rm Ca\,\textsc{ii}}$ likely reflects its strong preference for dense and weakly ionized environments. \cite{liang16} show that ion column density profiles follow $N(D)=N_0 \exp(-D/h_s)$, with $h_s \propto E_{\rm ion}^{0.74}$, where $E_{\rm ion}$ denotes the ionization potential. Thus, the low ionization potential of Ca\,\textsc{ii} ($\sim 12\,\mathrm{eV}$; \citealt{kramida2024}) limits its survival to the most dense regions of galaxy halos at small radii, naturally producing a steeper radial decline.
Moreover, compared to Mg\,\textsc{ii}, Ca\,\textsc{ii} is more easily depleted onto dust, which may further limit its transport into the more extended CGM~\citep{richter11}. 

We also constructed a foreground sample centered on QSOs, which typically reside in significantly more massive dark matter halos ($M_{\rm halo} \sim 10^{12.5} \rm {M_\odot}$) compared to the ELG and BGS samples~\citep{zhao13}. Due to the relatively high redshifts of the QSOs, we are able to examine C\,\textsc{iv} and Mg\,\textsc{ii} absorption in fully overlapping sightlines.  As shown in the top panel of Figure~\ref{fig:EW_spec_QSO}, similar to the results for ELGs, the absorption strengths of both Mg\,\textsc{ii} and C\,\textsc{iv} follow tight power-law distributions, with Mg\,\textsc{ii} exhibiting a steeper slope. This indicates that the trend of cooler gas being progressively more centrally concentrated holds across different galaxy types, and is also consistent with the findings of \cite{chen25b} on scales of $\sim1$\,Mpc. In contrast to the BGS sample, the TNG50 predictions for QSO-host systems yield Mg\,\textsc{ii} and C\,\textsc{iv} equivalent widths that are systematically higher than the observations by a factor of $\sim2$, while the measured $\alpha$ remains in good agreement ($-1.20$ for Mg\,\textsc{ii} and $-0.79$ for C\,\textsc{iv}), indicating its success in capturing the radial profiles in massive halos.

Notably, even for the same absorption line, the $\alpha$ can vary significantly across different samples. For Mg\,\textsc{ii}, for instance, the gas is most centrally concentrated around ELGs, followed by BGSs, and least concentrated around QSOs. QSO host halos typically reside in more massive dark matter halos, exhibit higher virial temperatures, and are exposed to stronger AGN radiation fields~\citep{Cappelluti12,Timlin18,Aird21}. As a consequence, the centrally concentrated structure of cool gas is more easily disrupted, while high-ionization species become more spatially extended due to the hot halo environment. This also reflects that host galaxy properties, such as stellar mass and redshift, can significantly affect the radial profiles of $W_i$, and these effects will be further discussed in Section~\ref{sec:discussion}. This motivates our effort to use consistent sightlines for comparisons, or at least ensure similar foreground samples.


\subsection{$W$ Ratio of Different Ions}\label{subsec:ratio}

\begin{figure*}
    \centering
    \includegraphics[width=2.05\columnwidth]{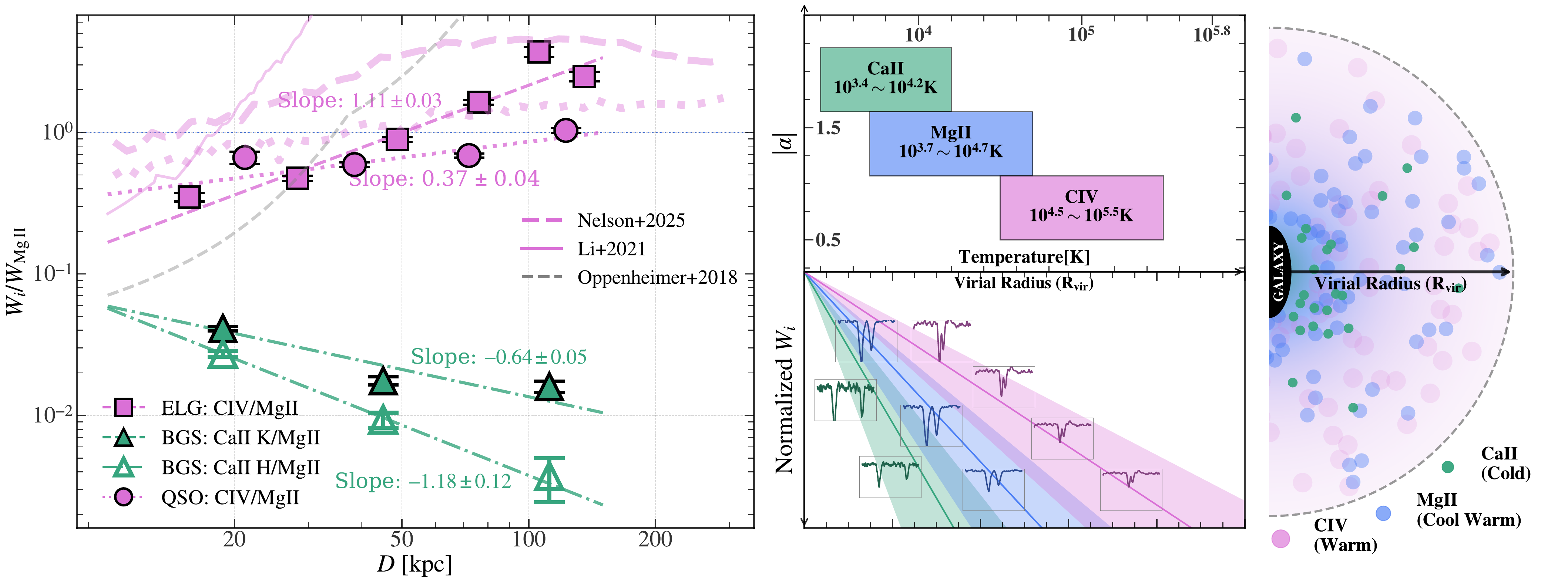}
    \caption{Left panel: radial dependence of the rest-frame equivalent-width ratios relative to Mg\,\textsc{ii} as a function of impact parameter. The symbols denote ratios derived from Gaussian fits to the stacked spectra: pink squares (ELGs) and circles (foreground quasar host galaxies) represent C\,\textsc{iv}/Mg\,\textsc{ii}, while green triangles indicate Ca\,\textsc{ii}/Mg\,\textsc{ii} in BGS halos (filled for the K transition and open for the H transition). The curves show the best-fitting power-law relations (dashed for ELGs, dotted for QSOs, and dash-dotted for BGS), with the corresponding slope values labeled in the panel. The thick pink curves present the $W_{\rm CIV}/W_{\rm MgII}$ measurements from SALSA~\citep{nelson2025}, including ELGs (dashed) and QSOs (dotted), while the thinner solid pink curve shows the prediction from the FIRE simulations (\citealt{li21}; see also Figure~\ref{fig:EW_spec_ELG}\,\&\,\ref{fig:EW_spec_QSO}). The gray dashed curve indicates the column density ratio of C\,\textsc{iv} to Mg\,\textsc{ii} ($N_{\rm CIV}/N_{\rm MgII}$) from the EAGLE zoom simulations ($10^{11.7} < M_{\rm halo} < 10^{12.3}\,\rm{M_\odot}$; \citealt{oppenheimer18}). Middle upper panel: schematic relation between gas temperature traced by different ionic species and the $|\alpha|$ of their radial profiles. Middle lower panel: radial profiles of normalized $W_i$ for different ions, together with representative absorption spectra, illustrating the relative central concentration of different ionic species in a qualitative manner. Right panel: Schematic spatial distribution of multiphase gas, including the cold phase (Ca\,\textsc{ii}, green), cool–warm phase (Mg\,\textsc{ii}, blue), and warm phase (C\,\textsc{iv}, pink), together with a diffuse halo background representing the surrounding warm/hot gaseous component. The cold gas is clumpy and centrally concentrated, whereas the warm gas is more diffuse and spatially extended. We adopt clumps of different characteristic sizes to illustrate this multiphase nature of the CGM, motivated by the fact that gas in the CGM is commonly distributed in discrete cloud-like structures rather than as a continuous medium~\citep{Vedantham2019,liang20,Sameer24}. In this picture, the colder phase tends to fragment into smaller, denser clumps, while being embedded within a more extended warm/hot medium~\citep{McCourt2018,sparre19,lan25}. }
    \label{fig_ratio}
\end{figure*}


The left panel of Figure~\ref{fig_ratio} shows Mg\,\textsc{ii}-related equivalent-width ratios ($W_i / W_{\rm Mg\,II}$) as a function of impact parameter, revealing the underlying phase structure. For the high-redshift samples (ELGs and QSOs), the C\,\textsc{iv}/Mg\,\textsc{ii} ratio increases monotonically with impact parameter, indicating that the outer halo is increasingly dominated by the warm, highly ionized gas phase. However, the steepness of this trend differs significantly between the two populations. The ratio around ELGs exhibits a rapid rise with a power-law index of $1.11 \pm 0.03$, whereas the profile around QSOs is much flatter ($0.37 \pm 0.04$). 

A similar increasing trend of C\,\textsc{iv}/Mg\,\textsc{ii} with $D$ is also seen in simulations, albeit with variations in amplitude and slope. This difference may partly arise from the use of median equivalent widths~\citep[e.g.,][]{oppenheimer18,li21} versus mean-based equivalent widths~\citep{nelson2025} (further discussion provided in Appendix~\ref{sec:appenA}). In addition, the non-linear relationship between $W_i$ and column density $N_i$ may also contribute to differences between \cite[e.g.,][]{oppenheimer18} and other studies. \cite{lan25} interpret the increasing $W_{\rm CIV} / W_{\rm Mg II}$ with impact parameter within a multiphase CGM framework, where different ionization phases are not strictly co-spatial but exhibit distinct radial distributions and covering fractions. 
As the Mg\,\textsc{ii}-bearing phase declines more rapidly with radius than the C\,\textsc{iv}-bearing phase, the resulting line-of-sight mixtures naturally produce an increasing C\,\textsc{iv}/Mg\,\textsc{ii} ratio toward larger galactocentric distances.

The difference in the $\alpha$ also reflects variations in the strength of the radial thermodynamic (or ionization) gradient within their host halos, with ELGs showing a more pronounced transition from cooler inner regions to a warmer, more highly ionized outer CGM compared to QSO host halos. In this picture, QSO halos maintain a relatively high ionization state throughout their volume, with the cool and warm gas phases more uniformly mixed, whereas ELG halos exhibit a clear ionization stratification, transitioning from a cool-gas-rich core to a warm-gas-dominated envelope. In contrast, the low-redshift BGS sample shows an opposite trend for the cold-to-cool gas ratio: the Ca\,\textsc{ii}/Mg\,\textsc{ii} ratio decreases with a negative slope ($\sim -0.6$ to $-1.2$), confirming that the cold gas phase traced by Ca\,\textsc{ii} is strongly confined to the inner regions, relative to the more spatially extended Mg\,\textsc{ii}-traced cool gas.


By sequentially comparing different tracers within matched samples (Ca\,\textsc{ii} vs. Mg\,\textsc{ii}, Mg\,\textsc{ii} vs. C\,\textsc{iv}), we have mapped the spatial distribution patterns of distinct gas phases in the CGM. As illustrated in the right panel of Figure~\ref{fig_ratio}, we present a schematic summary of the $\alpha$–temperature relation for the three ions, their corresponding equivalent-width radial profiles, as well as a cartoon representation of their two-dimensional spatial distributions. Here, the CGM shows a clear multiphase structure: the coldest gas (traced by Ca\,\textsc{ii}) is confined to the innermost regions, the cool gas (traced by Mg\,\textsc{ii}) extends farther out, and the warm gas (traced by C\,\textsc{iv}) forms the most extended component. Although the absolute radial extent and absorption strength of each phase depend on host galaxy properties, such as stellar and halo mass, the relative central concentration remains unchanged across different galaxies: cold gas is the most concentrated, followed by cool gas, and then warm gas~\citep{liang16,oppenheimer18}. 

\subsection{Dependence on Galaxy Properties} 

\begin{table*}[ht]
\centering
\caption{Summary of foreground samples used in the stellar mass analysis within $D<150\,$kpc (the left panel of Figure~\ref{fig_mass_denpendence}).}
\label{tab:sample6}

\begin{tabular}{llccc}
\toprule

Sample & $z$ selection & $\log(M_*/M_\odot)$ bin & $N_{\rm pair}$ & Used for \\
\midrule

\multirow{3}{*}{ELG}
& \multirow{3}{*}{$z > 1.3$}
    & $[9.2, 9.8]$ & 1400 & \multirow{3}{*}{Mg\,\textsc{ii}$+$C\,\textsc{iv}} \\
&
    & $[9.8, 10.2]$ & 3014 & \\
&
    & $[10.2, 11.0]$ & 1601 & \\

\midrule

\multirow{6}{*}{BGS}

& \multirow{3}{*}{$z > 0.3$}
    & $[9.4, 10.8]$ & 47128 & \multirow{3}{*}{Mg\,\textsc{ii}} \\
&
    & $[10.8, 11.1]$ & 42160 & \\
&
    & $[11,1, 12.0]$ & 29706 & \\

\cmidrule(lr){2-5}

& \multirow{3}{*}{$z > 0.1$}
    & $[9.4, 10.8]$ & 290636 & \multirow{3}{*}{Ca\,\textsc{ii}} \\
&
    & $[10.8, 11.1]$ & 71284 & \\
&
    & $[11.1, 12.0]$ & 62171 & \\

\bottomrule
\end{tabular}
\end{table*}

\label{sec:discussion}

\begin{figure*}
    \centering
    \includegraphics[width=1.8\columnwidth]{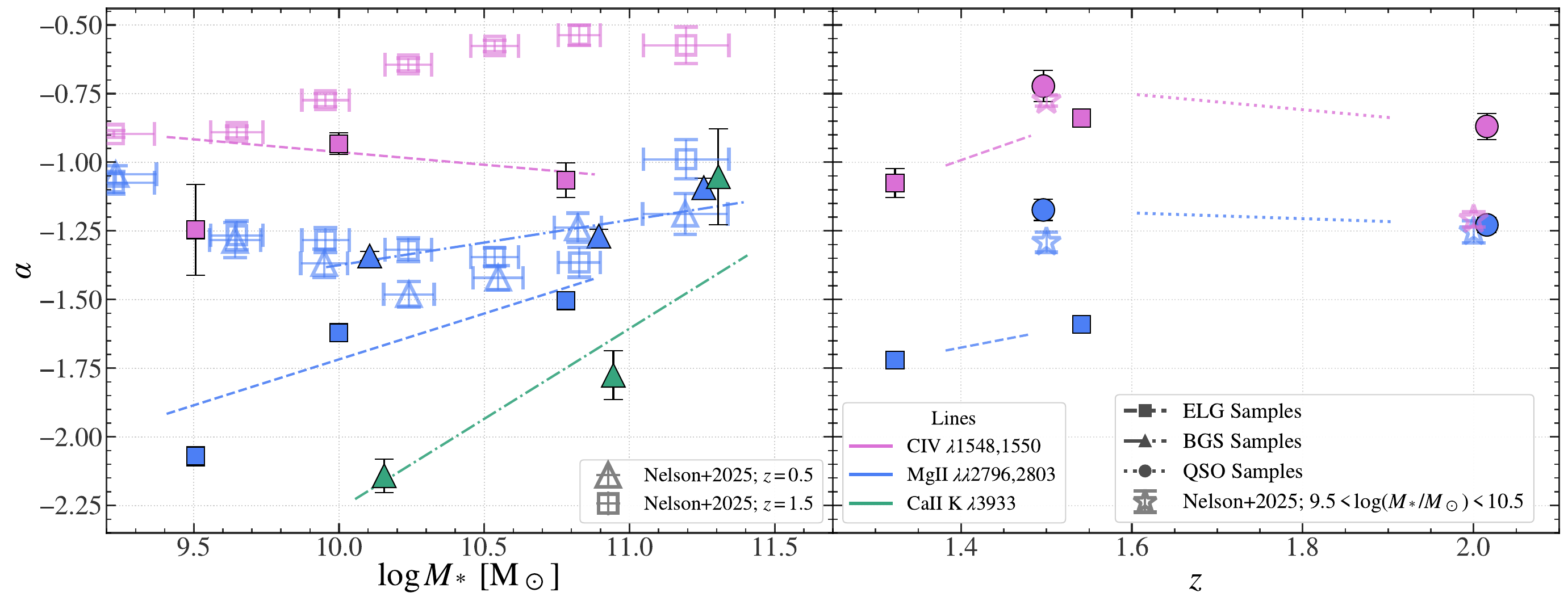}
    \caption{Left panel: dependence of the radial power-law index $\alpha$ on stellar mass. Symbols distinguish different ionic species (pink: C\,\textsc{iv}; blue: Mg\,\textsc{ii}; green: Ca\,\textsc{ii} K) and host galaxy populations (squares: ELGs; triangles: BGSs). The curves represent the best-fit linear trends, shown as dashed lines for ELGs and dash-dotted lines for BGSs. Right panel: redshift evolution of the radial power-law index $\alpha$ as a function of $z$. Data points are color-coded by ionic species (pink: C\,\textsc{iv}; blue: Mg\,\textsc{ii}) and distinguished by host galaxy type (squares: ELGs; circles: foreground QSOs). The dashed and dotted lines indicate the best-fit evolutionary trends for ELGs and QSOs, respectively. Error bars indicate 1-$\sigma$ uncertainties or the 1-$\sigma$ scatter of the mass distribution. Open symbols are taken from SALSA~\citep{nelson2025}, including measurements at $z=0.5$ (triangles) and $z=1.5$ (squares), as well as a stellar-mass-selected sample with $9.5 < \log (M_*/\rm {M_\odot}) < 10.5$ (pentagons), whose mass distribution is comparable to that of our ELG sample.}
    \label{fig_mass_denpendence}
\end{figure*}

To better understand the variations in the power-law index of the same absorption line among different foregrounds, we re-bin the foreground galaxies according to redshift and host galaxy mass, and measure their corresponding $\alpha$ values. Since QSOs lack host galaxy mass information, no such binning is applied to them. Similarly, for the BGS sample, the Mg\,\textsc{ii} redshift bins are too narrow to allow meaningful subdivision.

We first divide the ELG and BGS samples into three $M_*$ bins each, with the mean stellar masses spanning from $10^{9.5}$ to $10^{11.5}\,\rm M_\odot$. The selection of these subsamples and the corresponding number of quasar–galaxy pairs are summarized in Table~\ref{tab:sample6}. As shown in the left panel of Figure~\ref{fig_mass_denpendence}, for both Mg\,\textsc{ii} and Ca\,\textsc{ii}, the radial profiles clearly flatten with increasing host galaxy mass, whereas the C\,\textsc{iv} profile remains largely unchanged, consistently exhibiting a relatively shallow slope. At the high-mass end ($\log M_*/M_{\odot} > 11.0$), the $\alpha$ of the $W_i$ for the three ions tend to converge, suggesting a reduced distinction in the distribution of the multi-phase medium in massive halos. We further demonstrate in Appendix~\ref{sec:appenB}, by performing the analysis in bins of $R_{\rm vir}$, that this trend is primarily driven by stellar mass rather than differences in halo scale. This suggests that the observed dependence cannot be simply attributed to variations in virial radius among galaxies of different masses, but instead reflects an intrinsic connection between the CGM properties and the stellar mass of the host galaxies.

This change is primarily driven by the decrease of $W_{\rm Mg\,II}$ and $W_{\rm Ca\,II}$ in the inner regions, indicating that massive halos are experiencing heating effects that propagate from the inside out.
A plausible explanation is that massive halos undergo mass-dependent thermal evolution~\citep{Rees1977,White1978}: as halo mass increases, higher virial temperatures and more stable virial shocks~\citep[e.g.,][]{Dekel06} can support extended hot gaseous atmospheres, suppressing the survival of cold and low-ionization gas in the inner CGM. Such shock-heated halos are expected to become increasingly important above the characteristic halo mass where stable hot halos form, naturally reducing the distinction between different gas phases in massive systems.  Additional heating processes, including AGN feedback~\citep{bower06, croton06,gaspari12, Mo2024, Yao2024, LiHao2025, ChenYangyao2025, Jiang2025}, may further enhance this effect by injecting energy into the surrounding medium and maintaining hot halo conditions. The coldest gas traced by Ca\,\textsc{ii} is particularly affected, with its $\alpha$ evolving from $\sim -2$ to $\sim -1$, whereas C\,\textsc{iv}, already tracing relatively warm gas, is expected to be less sensitive to these heating processes.

In the TNG50 simulations, however, a different mass dependence is observed. The slope of the C\,\textsc{iv} profile increases steadily with stellar mass, while Mg\,\textsc{ii} is in good agreement with the observations at $\log M_*/M_{\odot} \gtrsim 10$, but shows an upturn at the low-mass end that is not seen in the data.
This discrepancy likely reflects differences in how feedback redistributes metals and gas across halo scales in the simulations~\citep{oppenheimer06,ford13,suresh2015}. In low-mass halos, stellar feedback in TNG50 appears to efficiently transport Mg\,\textsc{ii}-bearing cold gas to larger radii, leading to an artificially flattened or even inverted radial gradient. As a result, the inner halo is relatively depleted in Mg\,\textsc{ii}, while the outer regions are enhanced due to outflow-driven redistribution. 

In more massive halos, the evolution is instead governed by the growing importance of virial heating and a more volume-filling warm/hot circumgalactic medium~\citep{White1978,Dekel06,croton06}. In this regime, C\,\textsc{iv}-traced gas becomes increasingly extended and less centrally concentrated, leading to the observed mass-dependent steepening of its profile. The differing responses of Mg\,\textsc{ii} and C\,\textsc{iv} thus reflect their sensitivity to distinct CGM phases: cold clumps that are strongly affected by feedback-driven transport versus warm gas regulated by halo-scale thermal structure.

For the ELG and QSO samples, we further divide the sightlines with fully overlapping C\,\textsc{iv} and Mg\,\textsc{ii} coverage into two bins each, as shown in the right panel of Figure~\ref{fig_mass_denpendence}. From the behavior of QSOs, the redshift appears to have only a minor impact on the slopes of $W_{\rm Mg\,II}$ and $W_{\rm C\,IV}$, suggesting that the large-scale distribution of both cool and warm gas around QSO host halos remains relatively stable over the redshift range probed. This stability may reflect the dominance of the halo gravitational potential and long-term feedback processes, which set the large-scale CGM structure and mitigate the influence of cosmic time evolution. For ELGs, although there is a clear trend of the slopes flattening with increasing redshift, much of this effect can be attributed to the concomitant increase in galaxy stellar mass with redshift. Therefore, redshift likely plays only a secondary role in shaping the radial distribution of the gas~\citep{lan20,wu2024,chen25}.

The TNG50 simulation generally reproduces the observed $\alpha$ at redshifts $z = 1.5$ and $z = 2$ reasonably well. However, at $z = 2$, the $\alpha$ value for C\,\textsc{iv} is lower than expected, leading to a trend in which $\alpha$ flattens toward lower redshift. This discrepancy is likely driven by the lower stellar masses of simulated galaxies compared to observed QSO hosts, combined with the positive dependence of the $W_{\rm C\,IV}$–$\alpha$ relation on stellar mass.




\section{Summary} \label{sec:summary}

This study presents a systematic analysis of the multiphase CGM around galaxies and quasars utilizing the DESI Y1 dataset. We focus on three absorption lines, including Ca\,\textsc{ii}\,$\lambda\lambda$3934,\,3969, Mg\,\textsc{ii}\,$\lambda\lambda$2796,\,2803, and C\,\textsc{iv}\,$\lambda\lambda$1548,\,1550, across a wide redshift range ($0.1 < z < 4$) and impact parameters up to $\sim 150$ kpc. These three classical absorption lines trace gas spanning a temperature range of $T \sim 10^{3.5}$–$10^{5.5}$ K, probing the cold, cool, and warm phases, thus serving as key diagnostics of the multiphase CGM.
By combining DESI observations with predictions from cosmological simulations (e.g., TNG50, FIRE, and EAGLE), we explore how their spatial distributions differ and how they relate to the properties of the host galaxies. Our main findings are summarized as follows:

\textit{Spatial Distribution.} Within comparable foreground samples, the radial profile of $W_{\rm Mg\,II}$ is more centrally concentrated than that of C\,\textsc{iv}  (Figure~\ref{fig:EW_spec_ELG}\,\&\,\ref{fig:EW_spec_QSO}), while the Ca\,\textsc{ii} profile is even more centrally concentrated than Mg\,\textsc{ii} (Figure~\ref{fig:EW_spec_BGS}). This suggests that in the multiphase CGM, progressively colder components are increasingly confined to the dense central regions, whereas the warmer phases exhibit a more extended and stable distribution. This phase-dependent stratification is also reproduced in TNG50~\citep{nelson2025} and FIRE~\citep{li21} simulations, which suggests that it arises from the combined effects of feedback-driven outflows, gas recycling, and halo cooling, with low-ionization gas confined to inner regions and high-ionization gas more extended.

\textit{Ratio Profiles.} For ELGs, the $W_{\rm C\,IV}/W_{\rm Mg\,II}$ ratio rises steeply from the inner to outer halo, indicating a rapid transition from a cool-gas-dominated core to a warm-gas-dominated envelope (Figure~\ref{fig_ratio}). In contrast, QSO halos show a much flatter ratio profile, suggesting that AGN feedback maintains a relatively uniform distribution of warm and cool gas phases throughout the halo. In the BGS sample, the rapidly declining $W_{\rm Ca\,II}/W_{\rm Mg\,II}$ ratio with distance reflects the strict confinement of the cold Ca\,\textsc{ii} phase to the dense central regions, regulated by both gas density and temperature.

\textit{Dependence on Galaxy Properties.}
$W_i \propto D^{\alpha}$, with the index $\alpha$ varying across foreground samples and likely driven primarily by the stellar mass of the host galaxies (Figure~\ref{fig_mass_denpendence}). This trend, particularly pronounced for the colder tracers, likely reflects AGN-driven heating that progressively rises with galaxy mass, disrupting the cold structures in the inner halo and thereby potentially explaining their quenching behavior. While simulations broadly reproduce the observed radial scaling relations, they exhibit differences in normalization and mass dependence, suggesting that the efficiency of feedback and the redistribution of cool gas remain key uncertainties in current models.

These absorption-line measurements provide a critical benchmark for cosmological simulations, offering new constraints on feedback processes and the multiphase structure of the CGM. As mentioned above, a more centrally concentrated distribution of cooler gas may result from the preferential transport of hot, metal-enriched gas into the outer CGM via high-velocity outflows~\citep{shen12}, while the colder or more dust-prone components remain confined to the inner regions~\citep{richter11}. With the advent of high-resolution imaging from space telescopes, such as Euclid~\citep{Euclid22,Euclid2024} and the Roman Telescope~\citep{romanwang2022}, this scenario could be tested through the anisotropy in the spatial distribution of multiphase gas. Meanwhile, other large cosmological surveys, such as DESI-II and Spec-5~\citep{Schlegel22}, will enable systematic statistical measurements of the CGM at higher redshifts, thereby advancing our understanding of galaxy evolution and the baryon cycling.


\section*{Acknowledgements}
EW thanks the support of the National Science Foundation of China (Nos. 12473008) and the Start-up Fund of the University of Science and Technology of China (No. KY2030000200). KW acknowledges support from the Science and Technologies Facilities Council (STFC) through grant ST/X001075/1. CL is supported by the Fundamental Research Funds for the Central Universities (WK2030250123).
This research used data obtained with the Dark Energy Spectroscopic Instrument (DESI). DESI construction and operations is managed by the Lawrence Berkeley National Laboratory. This material is based upon work supported by the U.S. Department of Energy, Office of Science, Office of High-Energy Physics, under Contract No. DE–AC02–05CH11231, and by the National Energy Research Scientific Computing Center, a DOE Office of Science User Facility under the same contract. Additional support for DESI was provided by the U.S. National Science Foundation (NSF), Division of Astronomical Sciences under Contract No. AST-0950945 to the NSF’s National Optical-Infrared Astronomy Research Laboratory; the Science and Technology Facilities Council of the United Kingdom; the Gordon and Betty Moore Foundation; the Heising-Simons Foundation; the French Alternative Energies and Atomic Energy Commission (CEA); the National Council of Humanities, Science and Technology of Mexico (CONAHCYT); the Ministry of Science and Innovation of Spain (MICINN), and by the DESI Member Institutions: www.desi.lbl.gov/collaborating-institutions. The DESI collaboration is honored to be permitted to conduct scientific research on I’oligam Du’ag (Kitt Peak), a mountain with particular significance to the Tohono O’odham Nation. Any opinions, findings, and conclusions or recommendations expressed in this material are those of the author(s) and do not necessarily reflect the views of the U.S. National Science Foundation, the U.S. Department of Energy, or any of the listed funding agencies.

\appendix
\section{Comparison of Individual and Stacked Absorption} \label{sec:appenA}
\renewcommand\thefigure{A}

\begin{figure*}
    \centering
    \includegraphics[width=1.9\columnwidth]{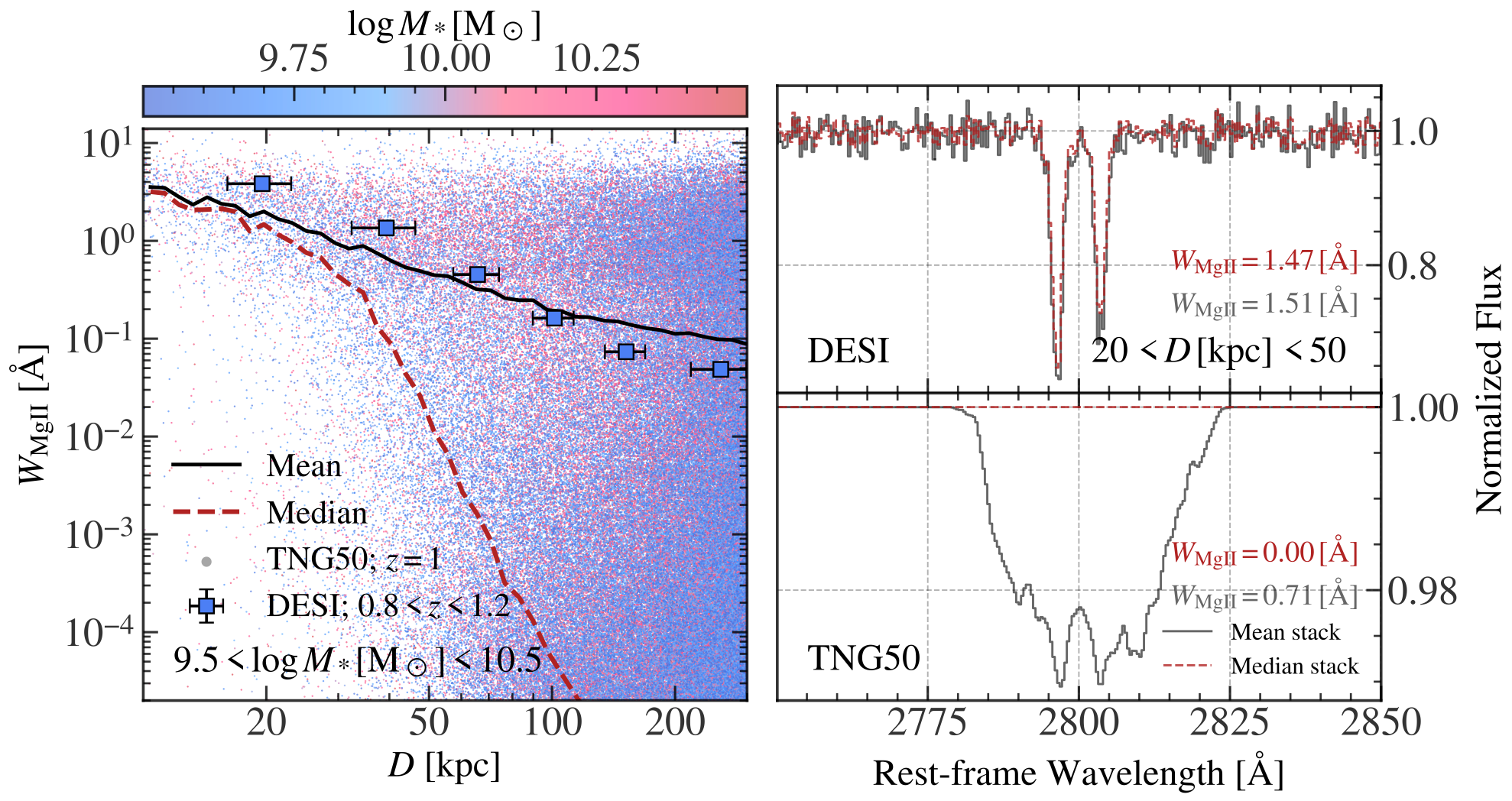}
    \caption{Comparison between individual absorbers in simulations and stacked observations.  
Left: Distribution of Mg\,\textsc{ii} $W_{\mathrm{MgII}}$ as a function of projected distance $D$ in the TNG50 simulation~\citep{nelson2025} for galaxies with $9.5 < \log (M_*/M_\odot) < 10.5$ at $z \sim 1$.  Individual absorbers are shown as points, color-coded by stellar mass. The black solid and red dashed curves denote the mean and median $W_{\mathrm{MgII}}$, respectively. Blue squares represent stacked measurements from DESI for galaxies at $0.8<z<1.2$ in the same stellar mass.  
Right: Stacked Mg\,\textsc{ii} absorption spectra at $20 < D < 50$ kpc. For DESI (top), the spectra are constructed using median stacking and a $3\sigma$-clipped mean. For TNG50 (bottom), only mean and median stacks are shown, without outlier rejection, as the synthetic spectra do not exhibit extreme flux outliers commonly present in observational data.}
    \label{test}
\end{figure*}

This work involves a comparison between observations~\citep[e.g.,][]{zhu13ca,ng2025} and simulations~\citep[e.g.,][]{li21,oppenheimer18}, as well as between composite (stacked) spectra~\citep[e.g.,][]{chen25} and individual absorbers~\citep[e.g.,][]{lan25,nelson2025}; particular caution is therefore required when performing such cross-scale comparisons. We compare our median-stacked results with the mean equivalent widths derived from individual absorbers and simulations, denoted as $W^{\rm stack}_{\rm median}$ and $W^{\rm ind}_{\rm mean}$, respectively. In practice, there is a substantial difference between the mean and median $W^{\rm ind}$ in both observational data and simulations. Such differences arise because the $W^{\rm ind}$ distribution is highly skewed, with a large fraction of nondetections (or zero absorption in simulations) and a small number of strong absorbers. Consequently, $W^{\rm ind}_{\rm mean}$ is dominated by rare high-$W$ systems, while $W^{\rm ind}_{\rm median}$ reflects the bulk of weak or undetected absorbers. In observations, nondetections are further limited by sensitivity, whereas in simulations they correspond to genuinely gas-poor sightlines.

The left panel of Figure~\ref{test} shows the distribution of individual absorbers in TNG50 from SALSA~\citep{nelson2025}, clearly illustrating the difference between $W^{\rm ind}_{\rm mean}$ and $W^{\rm ind}_{\rm median}$. At larger impact parameters, $W^{\rm ind}_{\rm median}$ drops to zero, whereas $W^{\rm ind}_{\rm mean}$ continues to follow a power-law radial profile that is broadly consistent with the observations. The right panel shows the spectral results. Interestingly, in the observations (DESI), $W^{\rm stack}_{\rm median}$ and $W^{\rm stack}_{\rm mean}$ are consistent, as expected since the median is a robust estimator of the mean that is less sensitive to outliers. In contrast, the median-stacked spectra in the simulations (TNG50) approach zero ($W^{\rm stack}_{\rm median} = 0$) because most sightlines exhibit zero equivalent width, while the remaining absorbers are distributed across different velocity components. 

Because the simulated spectra are close to ideal, $W^{\rm stack}_{\rm mean}$ measured from TNG50 is nearly identical to $W^{\rm ind}_{\rm mean}$. This equivalence can be understood analytically from the linearity of summation: since both stacking and equivalent-width integration are linear operations, the order of summation over sightlines and wavelength bins can be exchanged, i.e.
\[
\sum_i \sum_j \rm{flux}_i(\lambda_j)\,\Delta \lambda
=
\sum_j \sum_i \rm{flux}_i(\lambda_j)\,\Delta \lambda,
\]
which directly leads to $W^{\rm stack}_{\rm mean} = W^{\rm ind}_{\rm mean}$ in the absence of noise or nonlinear selection effects. Therefore, in this work it is reasonable to compare $W^{\rm ind}_{\rm mean}$ from simulations with $W^{\rm stack}_{\rm median}$ from observations. This is justified because in simulations $W^{\rm ind}_{\rm mean} = W^{\rm stack}_{\rm mean}$, while in observations $W^{\rm stack}_{\rm mean} \simeq W^{\rm stack}_{\rm median}$ owing to the robustness of median stacking against outliers. This correspondence can be equivalently understood as a comparison between simulations and observations both using a sigma-clipped mean stacking estimator.

In comparisons between observations, our results ($W^{\rm stack}_{\rm median}$) are in good agreement with \citealt{lan25} ($W^{\rm ind}_{\rm mean}$; see Figure~\ref{fig:EW_spec_ELG}), as their mean estimates also include nondetected sightlines. However, it is worth noting that these nondetected sightlines do not necessarily correspond to a complete absence of absorption; some relatively strong absorption may still be present but remains hidden within the noise of individual spectra. These weak absorption features only become detectable when a large number of spectra are stacked to improve the signal-to-noise ratio. Such weak absorption features are only revealed through stacking, which motivates the use of stacking-based methods to characterize the statistical properties of the CGM in this work.

\section{Halo Size Effect} \label{sec:appenB}
\renewcommand\thefigure{B}
The comparison in this paper using a fixed physical distance therefore inevitably mixes halo-mass and virial-radius effects. Massive galaxies have larger halos, and comparisons at fixed kpc may therefore make the profiles appear flatter or reduce the apparent phase distinction. Here, we discuss this effect in more detail. For the CGM of an individual galaxy, if $W \propto D^{\alpha}$, then it can also be expressed as $W \propto (D/R_{\rm vir})^{\alpha}$ up to a normalization factor. Therefore, if the redshift and stellar mass distributions of the sample are well controlled, the scatter in $R_{\rm vir}$ within each bin is small. In this case, binning in $D$ or in $R_{\rm vir}$ yields consistent results: $W \propto D^{\alpha}$ (equivalently $W \propto (D/\langle R_{\rm vir}\rangle)^{\alpha}$) is consistent with $W \propto (D/R_{\rm vir})^{\alpha}$, and the three forms yield approximately the same $\alpha$. However, in practice such control is rarely perfect, and the resulting scatter in $R_{\rm vir}$ within a given $D$ bin may introduce additional stacking-induced effects.

\begin{figure*}
    \centering
    \includegraphics[width=1.9\columnwidth]{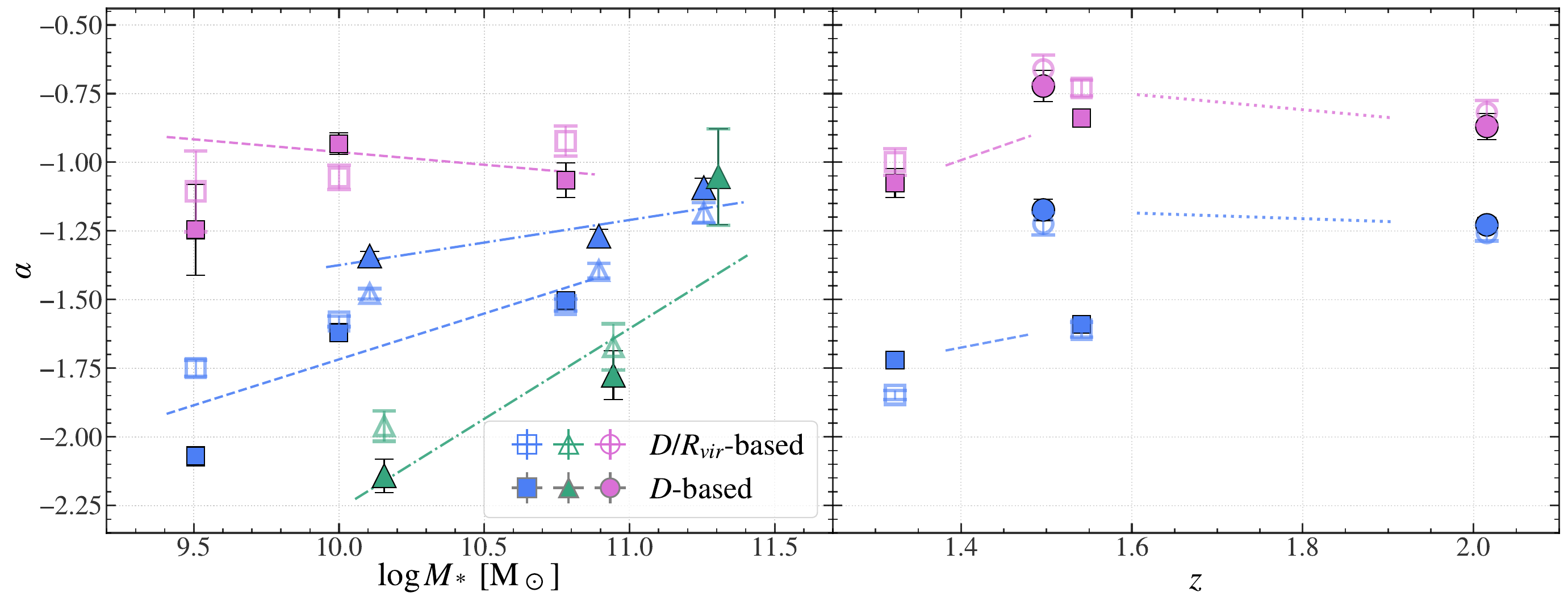}
    \caption{The same as Figure~6, but additionally showing $\alpha$ obtained using $D/R_{\rm vir}$ binning (open symbols). The halo-scaled radial bins are defined as $0.05\text{–}0.15$, $0.15\text{–}0.3$, $0.3\text{–}0.6$, and $0.6\text{–}1.0\,R_{\rm vir}$.}
    \label{appendix2}
\end{figure*}

In addition, measurements of $\alpha$ at a fixed physical scale effectively probe different relative locations within halos of different sizes. However, the Mg\,\textsc{ii}~\citep[e.g.,][]{zhu14,lan18,wu2024,chen25} and Ca\,\textsc{ii}~\citep[e.g.,][]{zhu13ca,ng2025} profiles show broadly consistent values of $\alpha$ within the one-halo regime, suggesting that comparisons across different relative halo regions remain physically meaningful. To quantify the halo size effect, we re-bin the sample shown in Figure~\ref{fig_mass_denpendence} in units of $R_{\rm vir}$ and obtain the corresponding $D/R_{\rm vir}$-based $\alpha$. As shown in Figure~\ref{appendix2}, the differences between the two binning schemes are not significant, suggesting that halo size effects do not impact the main conclusions of this work.

The virial radius of foreground galaxies (or QSOs) is calculated by the following relation~\cite{bryan98}: 
\begin{equation}
R_{\rm vir} = \left( \frac{3 M_{\rm halo}}{4\pi \Delta_{\rm c}(z)\,\rho_{\rm c}(z)} \right)^{1/3}.
\label{eq:Rv}
\end{equation}
For the QSO sample, we adopt a characteristic halo mass of $M_{\rm halo} \sim 10^{12.5}\,\rm M_\odot$~\citep{zhao13}. For ELGs and BGSs, we derive $M_{\rm halo}$ using the stellar-to-halo mass relation (SHMR) of \cite{moster13}, which can be written as:
\begin{equation}
\frac{M_*}{M_{\rm halo}}(M_{\rm halo}, z) = 2\,f_b\,\epsilon_{\mathrm{N}} \left[ \left( \frac{M_{\rm halo}}{M_1} \right)^{-\beta} + \left( \frac{M_{\rm halo}}{M_1} \right)^{\eta} \right]^{-1},
\label{eq:SHMR}
\end{equation}
where $f_b = 0.157$ is the universal baryon fraction, $\epsilon_{\mathrm{N}}$, $M_1$, $\beta$, and $\eta$ are adopted at $z=0.5$ for BGS and $z=1$ for ELGs, respectively. 

\bibliography{maix}
\bibliographystyle{aasjournal}

\end{document}